\documentclass[conference]{IEEEtran}
\IEEEoverridecommandlockouts
\usepackage[section]{placeins}

\usepackage{cite}
\usepackage{amsmath,amssymb,amsfonts}
\usepackage{algorithmic}
\usepackage{graphicx}
\usepackage{multirow}
\usepackage[utf8]{inputenc}
\usepackage{textcomp}
\usepackage{hhline}
\usepackage{booktabs}
\usepackage{tikz}
\usepackage{subcaption}
\usepackage{colortbl}
\usepackage{xcolor}
\usepackage{hyperref}
\usepackage{amsmath}

\def\BibTeX{{\rm B\kern-.05em{\sc i\kern-.025em b}\kern-.08em
    T\kern-.1667em\lower.7ex\hbox{E}\kern-.125emX}}
\usepackage{cite}
\def\BibTeX{{\rm B\kern-.05em{\sc i\kern-.025em b}\kern-.08em
    T\kern-.1667em\lower.7ex\hbox{E}\kern-.125emX}}    
\begin{document}

\title{Complete Cross-triplet Loss in Label Space for Audio-visual Cross-modal Retrieval}


\author{
\IEEEauthorblockN{Donghuo Zeng}
\IEEEauthorblockA{\textit{KDDI Research, Inc.} \\
Saitama, Japan \\
do-zeng@kddi-research.jp
}
\and
\IEEEauthorblockN{Yanan Wang}
\IEEEauthorblockA{\textit{KDDI Research, Inc.} \\
Saitama, Japan \\
ya-wang@kddi-research.jp}

\and
\IEEEauthorblockN{Jianming Wu}
\IEEEauthorblockA{\textit{KDDI Research, Inc.} \\
Saitama, Japan \\
ji-wu@kddi-research.jp}

\and
\IEEEauthorblockN{Kazushi Ikeda}
\IEEEauthorblockA{\textit{KDDI Research, Inc.} \\
Saitama, Japan \\
kz-ikeda@kddi-research.jp}
}

\markboth{Journal of \LaTeX\ Class Files,~Vol.~6, No.~1, January~2007}%
{Shell \MakeLowercase{\textit{et al.}}: Bare Demo of IEEEtran.cls for Journals}

\maketitle
\thispagestyle{empty}

\begin{abstract}
The heterogeneity gap problem is the main challenge in cross-modal retrieval. Because cross-modal data (e.g. audio-visual) have different distributions and representations that cannot be directly compared. To bridge the gap between audio-visual modalities, we learn a common subspace for them by utilizing the intrinsic correlation in the natural synchronization of audio-visual data with the aid of annotated labels. TNN-C-CCA is the best audio-visual cross-modal retrieval (AV-CMR) model so far, but the model training is sensitive to hard negative samples when learning common subspace by applying triplet loss to predict the relative distance between inputs. In this paper, to reduce the interference of hard negative samples in representation learning, we propose a new AV-CMR model to optimize semantic features by directly predicting labels and then measuring the intrinsic correlation between audio-visual data using complete cross-triple loss. In particular, our model projects audio-visual features into label space by minimizing the distance between predicted label features after feature projection and ground label representations. Moreover, we adopt complete cross-triplet loss to optimize the predicted label features by leveraging the relationship between all possible similarity and dissimilarity semantic information across modalities. The extensive experimental results on two audio-visual double-checked datasets have shown an improvement of approximately 2.1\% in terms of average MAP over the current state-of-the-art method TNN-C-CCA for the AV-CMR task, which indicates the effectiveness of our proposed model.
\end{abstract}

%
\IEEEpeerreviewmaketitle
\section{Introduction}
Vision and hearing are two important senses that humans use to perceive their surroundings and understand the world by capturing their intertwined relationships. For instance, when we hear the sound of a baby crying in another room, we subconsciously imagine his crying face and feel anxious as we walk toward the baby. Therefore, the intrinsic correlations in the natural synchronization between audio and visual data accompany with events/labels that can be modeled as cross-modal retrieval between these two modalities, i.e., given an audio/visual as a query, the system can automatically retrieve the relevant data that have the same annotated events as the query from another modality.

The main challenge in AV-CMR task is the heterogeneous gap that makes it impossible to measure cross-modal data each other directly, due to the different distributions of audio and visual data. In recent years~\cite{zhu2021learning, chen2017deep, alwassel2020self, zheng2021adversarial, rahman2021tribert}, the correlations between audio and visual information are the key to solutions in this field. Previous studies~\cite{zeng9232663dh, yu2018category, zeng2020deep, zeng2018audio, zeng2019learning, tian2018audio, xu2020cross} focused on representation learning methods to bridge the heterogeneous gap. These methods exploit the cross-modal correlations between each other in a common feature subspace to generate new representations of audio-visual data so that the generated representations across modalities can be compared.

Traditional representation learning methods for cross-modal retrieval such as canonical correlation analysis (CCA) tried to find linear projections to generate common representations by maximizing the inter-modality pairwise-based correlation or classification accuracy. However, the correlation between audio-visual data in the real world is too complex to be fully learnt by using only linear projections. The great success of deep neural networks (DNNs) has led to the advancement of cross-modal learning in finding a common representation subspace. DNNs have been utilized to improve CCA-based methods~\cite{andrew2013deep, yu2018category} through learning complex nonlinear transformations of cross-modal data. The current state-of-the-art work~\cite{zeng2020deep}, a triplet neural network with cluster canonical correlation analysis (TNN-C-CCA) applies audio-visual loss with a triplet neural network to improve the inter-modality pairwise-based correlations in the above CCA-based methods. The TNN-C-CCA model is trained in two steps, the first step is to use the audio-visual shared label to increase the number of audio-visual pairwises, where samples with the same label will be clustered during the training. The second step solves the issue where some samples were incorrectly clustered in the first step by using triplet loss method. However, the triplet loss is sensitive to those hard negative samples that will fail to capture semantic similarity during the training. Because the classification information (labels of samples) is not directly applied in the deep neural network training but only for preprocessing of the input data structure in the first step and the construction of input data for optimization of neural networks with triple loss. 

In the above cross-modal retrieval models, the classification information is underutilized in cross-modal representation learning. Therefore, we propose a novel AV-CMR model, which finds an effective common subspace based on fully exploiting the classification information and measuring the intrinsic correlation of audio-visual data.
Specifically, our proposed model projects audio-visual features into the label space by minimizing the distance between predicted label features after feature projection and ground label representations. Moreover, we adopt cross-triplet loss to optimize the predicted label features by leveraging the relationship between all possible similarity and dissimilarity semantic information between modalities.

In summary, this work makes the following contributions:
\begin{itemize}
    \item Heterogeneous data can be effectively represented by the common representation in the label space by fully mining semantic information with an optimized processing method (cross-triplet loss) in an end-to-end manner. 
    \item A linear classifier-based method is applied to learn predicted label features of audio-visual data. We utilize the label space that consists of predicted label features and ground label representations, where the classification information is fully exploited during the model training.
    \item An optimized loss function in the label space is used to learn the natural intrinsic correlations between audio-visual data by pulling all the similar pairs together while pushing all the dissimilar pairs apart.
    \item Extensive experiments on two audio-visual double-checked datasets have been conducted. The results demonstrate that our method surpasses current state-of-the-art methods for the AV-CMR task, which indicates that adopting the cross-triplet loss not only improves on single cross-triplet loss but can also optimize the semantic information in the label space.
\end{itemize} 


\section{Related work}
\label{relatedwork}
In this section, we review the common subspace learning and ranking loss for cross-modal retrieval, which are related to our proposed model.
\subsection{Common Subspace for Cross-modal Retrieval}
Common subspace learning is a frequently adopted method to bridge the heterogeneous gap to achieve the cross-modal retrieval. CCA~\cite{hardoon2004canonical} is a typical statistical approach for finding basis vectors for two sets of variables by optimizing the correlation between the linear projections of the two sets on basis vectors in the common space. To extend the linear projections to a type of complex mapping way, KCCA~\cite{akaho2006kernel} introduces a "kernel trick" method to project the cross-modal data into a high-dimensional space. However, it is hard to decide what kind of kernel method to use. In pursuit of more complex and flexible projections of multimodal data, DCCA~\cite{andrew2013deep} adopts a deep learning mechanism to acquire the nonlinear projection of two sets. DCCA can be viewed as an extension of CCA. C-CCA~\cite{rasiwasia2014cluster} applies the label information to cluster data across modalities by establishing all the possible correspondences, then using CCA to learn the intra-cluster correlation that can push the points of different labels apart in the common subspace. C-DCCA~\cite{yu2018category} is derived from both C-CCA and DCCA, projecting the cross-modal data into the common subspace through deep neural network mapping so that the data points within a cluster are highly correlated. The TNN-C-CCA~\cite{zeng2020deep} model can be regarded as an improvement over the combination of C-CCA and C-DCCA. It uses the label information to consider both similar and dissimilar correlations by applying audio-visual ranking loss to improve the pairwise-based CCA methods. It can achieve the best result on the VEGAS dataset we plan to use in this work.

Except for the CCA-based methods that aim at learning the pairwise-based correlation across modalities in the common subspace, some recent studies~\cite{wang2017adversarial, jiang2017deep, zhen2019deep} utilize the label information to design ranking loss to improve the semantic information of multimodal data in common subspace. The ACMR~\cite{wang2017adversarial} model applies label information to learn semantic representations for each modality in the common subspace. The DCMH~\cite{jiang2017deep} model uses label information to learn the discriminative information between inter-modality samples in a common subspace. The DSCMR~\cite{zhen2019deep} model employs label information to learn discriminative representations in the common subspace by minimizing the distance between the label space and the common subspace.  

\begin{figure*}[h]
\centering
\includegraphics[width=0.95\textwidth]{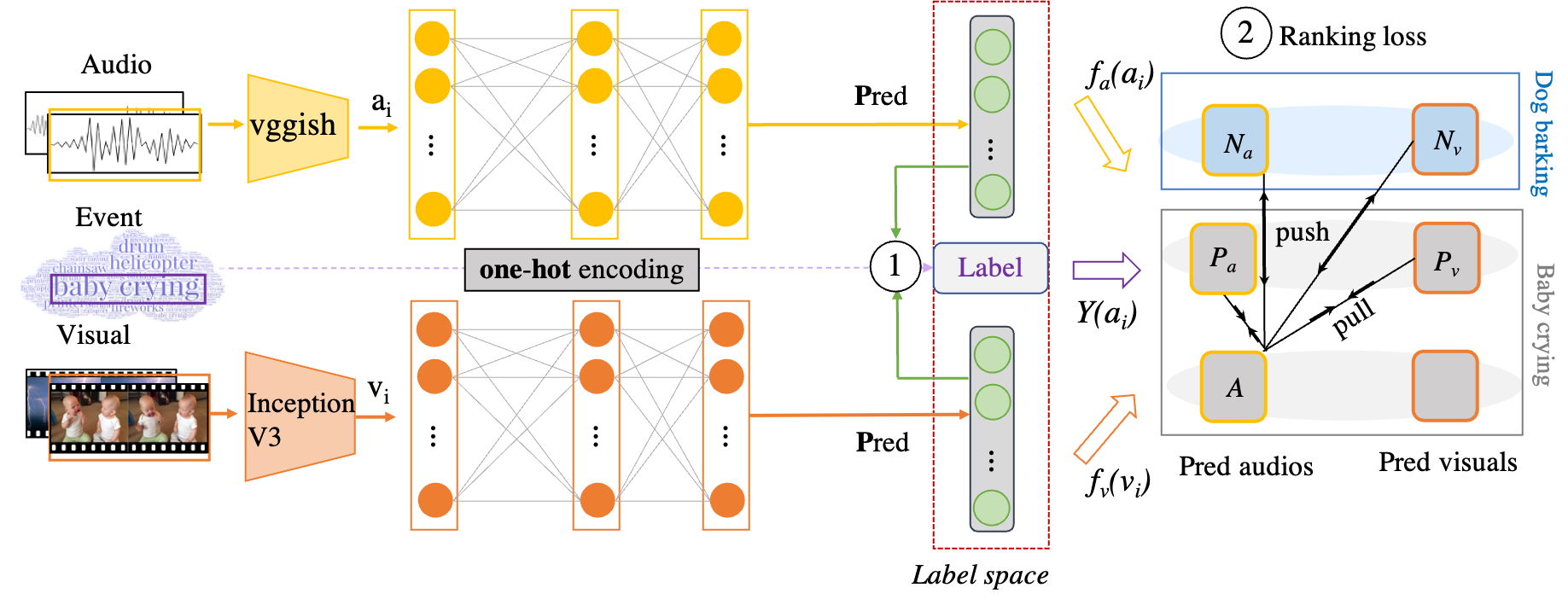}
\caption{Overview of our architecture. We project audio and visual representations $a_{i}$ and $v_{i}$ extracted by Vggish and InceptionV3 pretrained models into the shared label space by using a supervised learning function \textcircled{1} (seen in Equation [\ref{equ:dis}]) with labels (events) information, the label is represented as one-hot encoding. Then, applying a new \textcircled{2} complete cross-triplet loss as a ranking loss to learn correlations between predicted audio and visual label features, A, P, N are the predicted samples, where the similar data points ($A$ and $P$) are pulled together and the dissimilar data points ($A$ and $N$) are pushed apart. Where $A$ (with "Baby crying" event) is selected audio sample as an anchor, $P_{a}$ and $P_{v}$ (also with "Baby crying" event) are respectively positive audio and visual samples, $N_{a}$ and $N_{v}$ (with "Dog barking" event) are respectively negative audio and visual samples.}
\label{fig:arch}
\end{figure*}+

\subsection{Ranking Loss for Cross-modal Retrieval}
The objective of the ranking loss function in cross-modal retrieval is to predict relative distances between inputs from inter-modalities. The most common ranking loss function includes contrastive loss and triplet loss, which are applied in many different applications with similar formulations with minor variations. The XMC-GAN~\cite{zhang2021cross} model utilizes multiple contrastive losses to capture inter-modality and intra-modality correspondences to maximize the mutual information across modalities.
Triplet loss addresses an internal limitation in contrastive losses push force. The CCAL~\cite{dorfer2018end} model adopts a pairwise ranking loss (triplet loss) to optimize the projection of deep CCA on the output of the last layer. It improves triplet loss optimization by adding a CCA projection layer between the dual deep neural network and the optimization target. The DCIL~\cite{zheng2020dual} model solves the less effective issue when triplet loss is used in cross-modal learning with heterogeneous features directly because it is difficult to select appropriate triplets at the beginning of training. The model introduces an instance loss to capture intra-modal data representation to improve the triplets selection.

In summary, previous methods applied label information to learn semantic information within modality or capture the correlation between modalities with CCA-based method or ranking loss in a common subspace. Nevertheless, no studies have explored learning correlations across modalities and capturing semantic information within a modality in a shared semantic subspace. Therefore, our proposed method introduces a label space, where the obtained semantic features of cross-modal data and the original label representations are located. Then, the learned semantic features are improved by applying the intrinsic correlation between audio and visual features to consider the similar and dissimilar semantic information.

\section{The Proposed Method}
\label{architecture}
\subsection{Problem Formulation}
We assume a collection of $n$ instances of audio-visual pairs, denotes as $\Phi = \{(a_{i}, v_{i})\}_{i=1}^{n}$, where $a_{i}$ is the input audio with 128-D and $v_{i}$ is the input visual with 1024-D of the $i$-th example. Each $(a_{i}, v_{i})$ pair is annotated with a double-check single label $Y_{i} \in R^{c}$, where $c$ is the number of labels. In cross-modal retrieval, given a sample with its label from one modality as a query, the system will rank all the samples in the database from another modality based on the similarity between the query and each sample. In the end, the similarity of the samples from the same label as query is larger than those samples from the different labels as query. Therefore, the relevant samples from another modality in the database can be returned to the query.

To ensure audio and visual can be directly measured, we project them into a shared subspace by mapping the projected features into label space, the overall projection functions is represented as $f_{a}(a_{i})$ and $f_{v}(v_{i})$ for audio and visual transforms, respectively. We aim at learning more effective transform functions, i.e., to make both $f_{a}(a_{i})$ and $f_{v}(v_{i})$, and the output distributions of them to be more semantically discriminative and well maintain the intrinsic correlation between both modalities.

\subsection{Label Space Learning}
Because common subspace learning focuses on capturing correlations between modalities and applying ranking loss to predict relative distances between projected features, this will affect the learning of projected features used to directly predict a label. The reason is that label information is not directly used to map projected features to labels; therefore, the semantic information of the final output representations depends on the projected relative distance rather than the direct label prediction. In this case, the semantic information is not fully mined.

To preserve the discrimination of samples from different labels after the feature projection through the deep neural network, we add a linear classifier layer connected on top to the audio and the visual sub-networks. This classifier takes projected features as input and generates a predicted ten-dimensional embedding label for each sample. The objective function to measure the discriminative semantic information in the label space is as follows.
\begin{equation}
    Loss_{lab} = \frac{1}{n}||f_{a}(a_{i})-Y(a_{i})||_{F} + \frac{1}{n}||f_{v}(v_{i})-Y(v_{i})||_{F}
    \label{equ:dis}
\end{equation}
where $||\cdot||_{F}$ denotes the Frobenius norm, $f(x)$ is the projected feature of the linear classifier in the label space. The $Y(\dots)$ is a one-hot encoding that represents the ground-truth events of the samples, and the events are annotated by double-checking in the audio and visual modalities.

\subsection{Cross-triplet Loss}
\begin{figure}[h]
\centering
\includegraphics[width=0.46\textwidth]{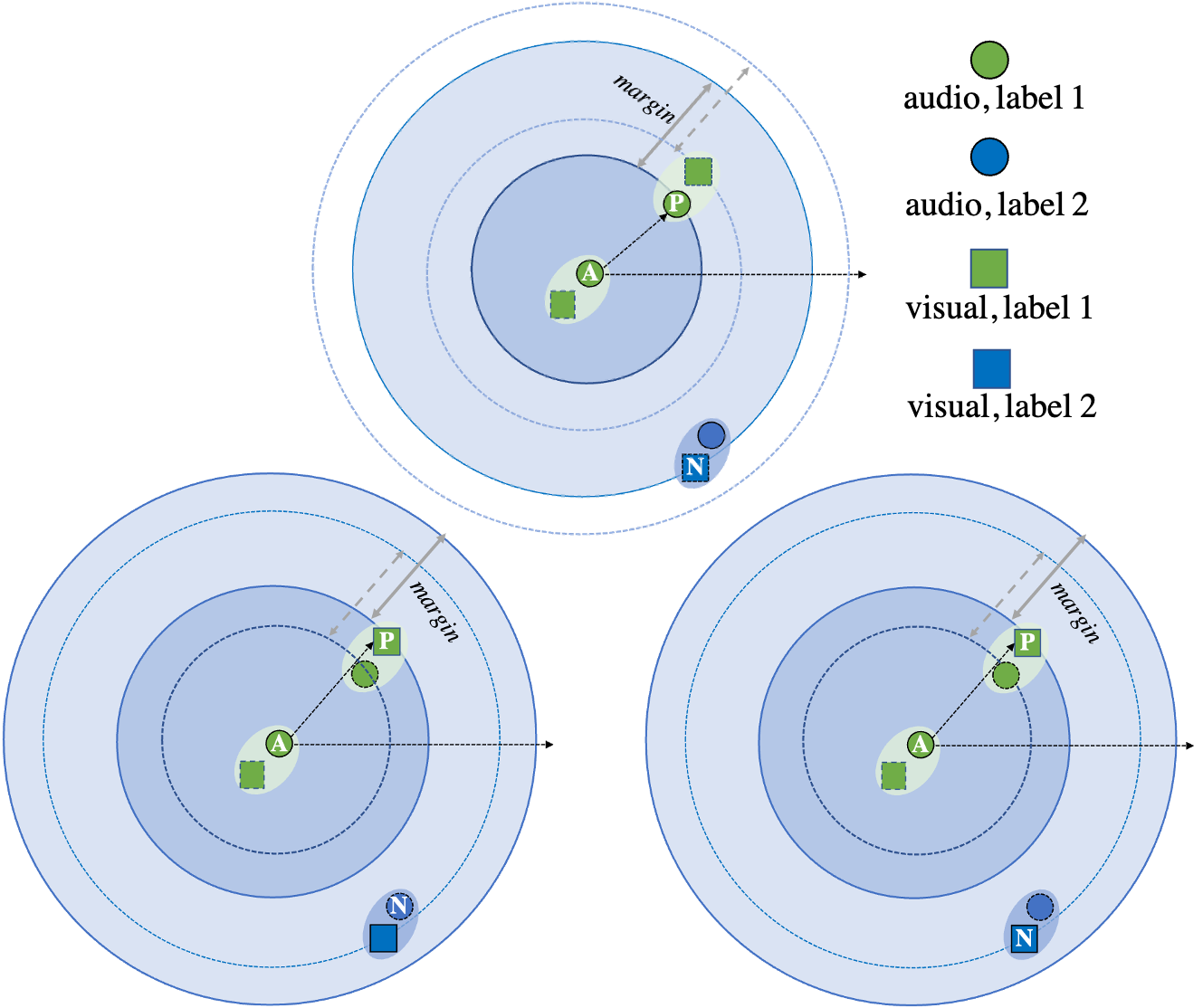}
\caption{Overview of three kinds of cross-triplet losses when selecting an audio sample with label 1 as an anchor (A) and setting a fixed margin. With the change of data modalities used as positive and negative samples, we list three kinds of cross-triplet for the audio as anchor. The top sub-figure selects an audio sample with label 1 as positive (P) and a visual sample with label 2 as negative (N), the bottom left sub-figure select a visual sample with label 1 as positive and audio sample with label 2 as negative, the bottom right sub-figure selects a visual sample label 1 as positive and another visual label 2 as negative.}
\label{fig:triplet}
\end{figure}
Learn label space by adding linear classifier is not enough for new representations generation, because it may overlook the useful information from the intrinsic commonality of cross-modal data. To explore the intrinsic correlation between audio and visual features and further extract cross-modal semantic features in combination with manually annotated events, all audio and visual pairs in the label space can be divided into internally related and unrelated at the semantic level. To perform this task, we adopt triplet loss~\cite{SchroffKP15} as the basis of our pairwise-based ranking loss.

Triplet loss is the most commonly used loss function for supervised similarity, which drives the distance between dissimilar pairs and any similar pair up to a certain margin, defined by: 
\begin{equation}
\begin{split}
    Loss = max(0, d(A, P) - d(A, N) + margin)
\label{equ:triplet}
\end{split}
\end{equation}

where $P$, i.e., \underline{P}ositive, is a sample in one modality having the same event/label as $A$, i.e., \underline{A}nchor. $N$, i.e., \underline{N}egative, is another sample in one modality that has an event/label different from $A$. Function $d$ is to measure the \underline{d}istance between the two samples. $margin$ is used to ensure a negative sample that is far apart from the positive samples.

In Fig.~\ref{fig:triplet}, when the anchor and positive are selected and the margin is set, the negative boundary is where the cross-triplet loss value of negative samples is zero, the loss value of negative samples located outside the boundary is greater than zero, and those inside samples which is zero, seen in the solid outer circle of the three subplots in Fig.~\ref{fig:triplet}. So the final cross-triplet loss depends on these negative samples that lie outside the negative boundary. We can find in Fig.~\ref{fig:triplet} that where the positive is set as audio or visual, this will impact the negative boundary and that will affect the effectiveness of the triplet losses. In this case, we consider both audio and visual as positive, as well as negative, and we ensure that the three components in a triplet (anchor, positive, negative) are not from the same modality. The final cross-triplet loss that can capture all the possible cross-triplets are summarized as the following formula.
\begin{equation}
\begin{split}
    \resizebox{0.95\hsize}{!}{$L_{cross} = max(0, \sum_{(i,j,k) \setminus \forall (i,j,k) \in \{A\mid V\}}^{\{A, V\}} d(A_{i}, P_{j}) -d(A_{i}, N_{k}) + margin)$}
    \label{equ:triplet_c}
\end{split}
\end{equation}
Where $i, j, k \in \{A, V\}$, i.e., \underline{A}udio and \underline{V}isual modalities. $(i,j,k)\setminus \forall (i,j,k)\in \{A\mid V\}$ denotes that the modality indexes $i,j,k$ are not from the same modality. Finally the above cross-triplet loss will be computed based on six combined cross-triplets. Our final objective function of the proposed method is as follows.
\begin{equation}
   L = Loss_{lab} + Loss_{cross}
    \label{equ:overall_formula}
\end{equation}
In the end, the final loss function is optimized by using the stochastic gradient descent (SGD) algorithm.

\section{Experiment}
\label{experiment}
\subsection{Dataset and Features}
To evaluate the effectiveness of our proposed model, we conduct experiments with two audio-visual double-checked datasets, i.e., VEGAS~\cite{zeng2020deep} and AVE~\cite{tian2018audio}. The VEGAS dataset has also been adopted in several other studies~\cite{zeng2018audio, zeng2019learning, zeng2020deep} of AV-CMR task. The VEGAS dataset is a subset of the Google AudioSet~\footnote{https://research.google.com/audioset/} that applies Amazon Mechanical Turk for data cleaning to ensure those manually annotated events appear in both audio and visual tracks. The events are human voices and natural sounds and each video contains only one event. The length of a video ranges from 2 to 10 seconds with an average of 7 seconds. In our experiments, we use 28,103 videos to evaluate our architecture, 80\% for training with the remaining 20\% for testing. The AVE dataset contains 15 events (clock, motorcycle, train horn, etc.) 
We select 1,955 videos with audio-visual tracks to evaluate our model,  1,564 videos as the training set, and 391 videos as the test set.

Following the work~\cite{zeng2020deep}, we use the same pretrained model extracted representations of audio-visual data. The audio is extracted by a pretrained Vggish model and the visual is extracted by a pretrained InceptionV3 model. Finally, each audio sample is represented by a 128-D vector and each visual sample is represented by a 1024-D vector.

\subsection{Evaluation Metric}
To leverage our proposed model, we employ mean average precision (MAP) to evaluate the AV-CMR performance on the VEGAS dataset, where a higher value of which indicates that the model's performance is better. In addition, to further evaluate the performance of our model, we apply precision-scope@K curves. In the process of model testing, given a query in one modality, the model will produce a ranking list in another modality for the query. We regard an item in the rank list with the same event as the query as correct.

\subsection{Experiment Setting}
In model training, the inputs of the audio and visual branches in the deep neural network are 128-D and 1024-D extracted features. In our experiments, we set the parameters of all models to be the same. The important parameters in the experiment settings are as follows.
\begin{itemize}
  \item The neural networks of both the audio and the visual branches have 3 hidden layers. The number of units per layer of audio and visual branches is 1024, 1024, 100. We apply the $Adam$~\cite{kingma2014adam} optimization algorithm for training our models.
  \item The training batch size is 512 and the test batch size is 64. The number of training epochs is 100.
  \item The learning rate is set to 0.0001.
  \item We set the margin for cross-triplet loss at 1.0.
\end{itemize}
     
We implement our model using PyTorch 1.12.0 and conduct training in the NVIDIA (42C) GeForce (P8) GPU with 10G memory running the Ubuntu 22.04 LTS Operation System.

\begin{table*}[h]
\caption{The MAP results of our model compared with other models on both audio-visual datasets (VEGAS and AVE datasets).}
\begin{center}
\scalebox{1.2}{%
\begin{tabular}{|c|c|c|c|c|c|c|}
    \hline
   \multirow{2}{*}{\textbf{Methods}} & \multicolumn{3}{c|}{\textbf{VEGAS Dataset}} & \multicolumn{3}{c|}{\textbf{AVE Dataset}} \\
    \cline{2-7}
  & audio$\rightarrow$visual
  & visual$\rightarrow$audio
  & Average
  & audio$\rightarrow$visual
  & visual$\rightarrow$audio
  & Average
   \tabularnewline 
\hline
   Random
    & 0.110 & 0.109 & 0.109 & 0.127 & 0.124 & 0.126\\ \hline
   CCA~\cite{hardoon2004canonical} 
    & 0.332 & 0.327  & 0.330  & 0.190 & 0.189 & 0.190 \\ \hline
   KCCA~\cite{akaho2006kernel}
    & 0.288 & 0.273  & 0.281  & 0.133 & 0.135 & 0.134\\ \hline
   DCCA~\cite{andrew2013deep}
    & 0.478 & 0.457  & 0.468 & 0.221 & 0.223 & 0.222 \\ \hline
   C-CCA~\cite{rasiwasia2014cluster}
    & 0.711 & 0.704  & 0.708 & 0.228 & 0.226 & 0.227 \\ \hline
   C-DCCA~\cite{yu2018category, zeng2018audio}
    & 0.722 & 0.716  & 0.719 & 0.230 & 0.227 & 0.229 \\ 
    \hline
   UGACH~\cite{ZhangPY18}
    & 0.182 & 0.179 & 0.181 & 0.165 & 0.159 & 0.162\\   \hline
   AGAH~\cite{GuGGLXW19}
    & 0.578 & 0.568 & 0.573 & 0.200 & 0.196 & 0.198\\ \hline
   UCAL~\cite{XLYSS17}
    & 0.446 & 0.436 & 0.441 & 0.153   & 0.150   & 0.152\\ \hline
   ACMR~\cite{wang2017adversarial} 
    & 0.465 & 0.442 & 0.454 &  0.162   & 0.159   & 0.161\\ \hline
   DSCMR~\cite{zhen2019deep}
    & 0.732 & 0.721 & 0.727 & 0.314 & 0.256 & 0.285\\ \hline
   TNN-C-CCA~\cite{zeng2020deep}
    & 0.751 & 0.738 & 0.745 & 0.253 & 0.258 & 0.256\\ \hline
   CLIP~\cite{pmlr_v139_radford21a}
    & 0.473 & 0.617 & 0.545 & 0.129 & 0.161 & 0.145\\ \hline
   BiC-Net~\cite{hou2021bicnet}
    & 0.680 & 0.653 & 0.667 & 0.188 & 0.187 & 0.188\\ \hline
   DCIL~\cite{zheng2020dual}
    & 0.726 & 0.722 & 0.724 & 0.244 & 0.213 & 0.228\\
    \hline
   \textit{\textbf{Our model} }
    & \textbf{0.766} & \textbf{0.765} & \textbf{0.766}
    & \textbf{0.328} & \textbf{0.267} & \textbf{0.298} \\
\hline
\end{tabular}}
\label{tab:result}
\end{center}
\end{table*}

\subsection{Comparison with State-of-the-art Models}
To verify the effectiveness of our proposed model, in the experiments, we compare our model with CCA-based methods and other eight other deep learning-based state-of-the-art methods in the experiments. These CCA-based methods include CCA, KCCA, DCCA, C-CCA, and C-DCCA, their central tasks are to find a common space so that the linear or nonlinear projections of two sets of variables can be computed and to ensure the correlations between them are mutually maximized. The TNN-C-CCA~\cite{zeng2020deep} model adopts C-CCA model and the outputs are mapped into a common subspace via a triplet neural network such that the data points belonging to the same label are highly correlated and the data points from different clusters are rarely correlated. The UGACH~\cite{ZhangPY18} model applied GAN to capture the potential manifold structure of cross-modal data by the k-nearest neighbors algorithm. The AGAH~\cite{GuGGLXW19} model utilizes an adversarial attention model to improve the discrimination of cross-modal representations. The UCAL~\cite{XLYSS17} and ACMR~\cite{wang2017adversarial} models respectively apply unsupervised and supervised adversarial learning for representation learning of cross-modal data, respectively. The DSCMR~\cite{zhen2019deep} is an advanced supervised cross-modal retrieval model and learns a very effective discriminative representation of cross-modal data. Three other recent models, CLIP~\cite{pmlr_v139_radford21a}, BiC-Net~\cite{hou2021bicnet}, and DCIL~\cite{zheng2020dual} models are employed in the experiments.

Table~\ref{tab:result} shows the MAP scores of our model in comparison with the other methods on two audio-visual datasets (VEGAS and AVE datasets). It can be observed that our model surpasses the previous best methods. The overall performance is summarized as follows.
\begin{itemize}
    \item Our model achieves an improvement of 1.5\%, and 2.7\% in terms of MAP for audio$\rightarrow$ visual and visual$\rightarrow$ audio respectively, and 2.1\% in terms of average MAP on the VEGAS dataset. We also used the AVE dataset to further leverage our model. It should be noted that our model can also achieve a competitive improvement in terms of the audio$\rightarrow$ visual and visual$\rightarrow$ audio retrieval process.
    \item These supervised learning models (C-CCA, C-DCCA, AGAH, DSCMR, ACMR, etc.) can achieve higher performance than unsupervised learning models (CCA, KCCA, DCCA, and UGACH) by using label information to train AV-CMR models, e.g. C-CCA outperforms CCA by a significant margin on the VEGAS dataset.
    \item The nonlinear projections in deep learning methods can improve the performance of traditional methods, e.g. C-DCCA and DCCA achieve a better performances than C-CCA and CCA, respectively.
    \item The ranking loss method can improve the performance of traditional CCA methods, e.g, using triplet loss in TNN-C-CCA can improve the C-CCA model in audio$\rightarrow$visual and vice versa on both datasets. 
    \item The cross-modal retrieval models applied to image-text data can achieve a good results, but they can not guarantee their performances on audio-visual data can also achieve high performance, e.g, the ACMR, CLIP, BiC-Net, and DCIL models outperforms almost all other models on some image-text data, but they all can not achieve the best performance on the audio-visual dataset. The best performing model, DCIL, still lags behind the current best model by 4.2\%.
\end{itemize}

\begin{figure*}[h]
\centering
\includegraphics[width=0.9\textwidth]{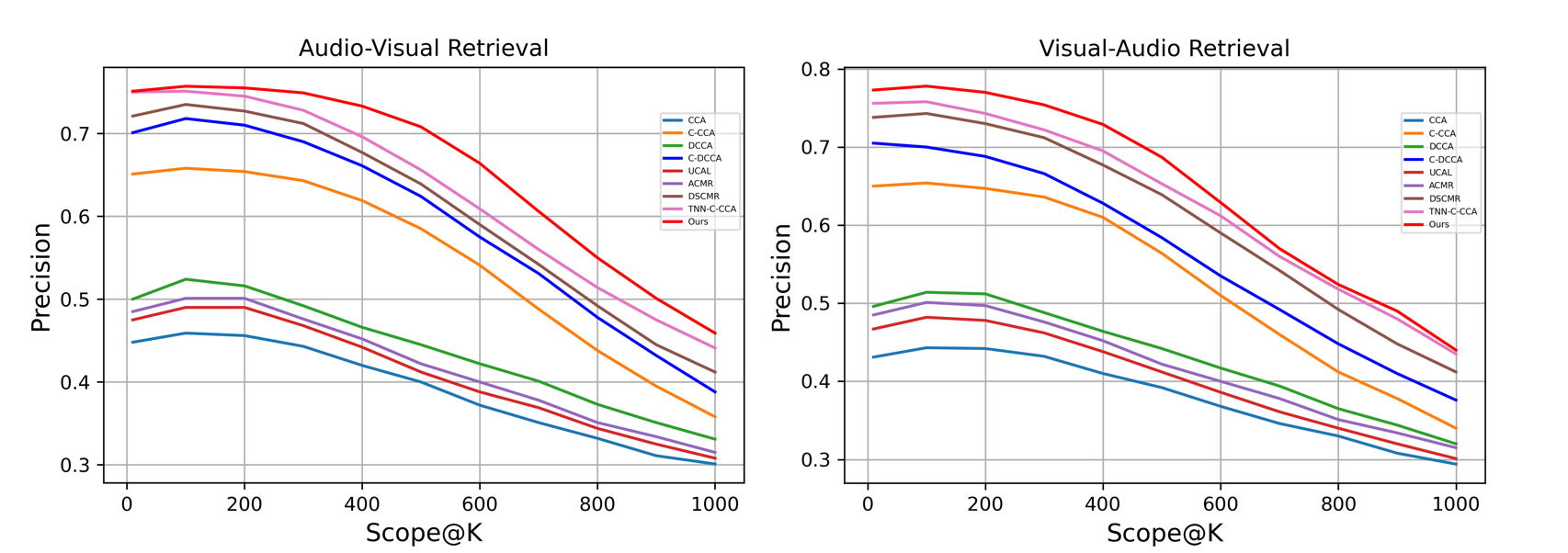}
\caption{Precision-scope curves of $audio\! \rightarrow \! visual$ and $visual\! \rightarrow \! audio$ retrieval experiments on VEGAS dataset, with K ranges from 10 to 1000.}
\label{lab:prec_rec}
\end{figure*}

Fig.~\ref{lab:prec_rec} shows the precision-scope@K curves of the different methods on VEGAS dataset, where K ranges from 1 to 1000. We depict the curves of all compared models in Table~\ref{tab:result}, our model achieves higher precision than the other models at the different levels of recall overall, which indicates that the matched samples of our model appear more often in the top of rank list and proves that our model convincingly achieves the best performance in AV-CMR in Table~\ref{tab:result}.

\begin{figure*}[t]
\centering
\includegraphics[width=0.9\textwidth]{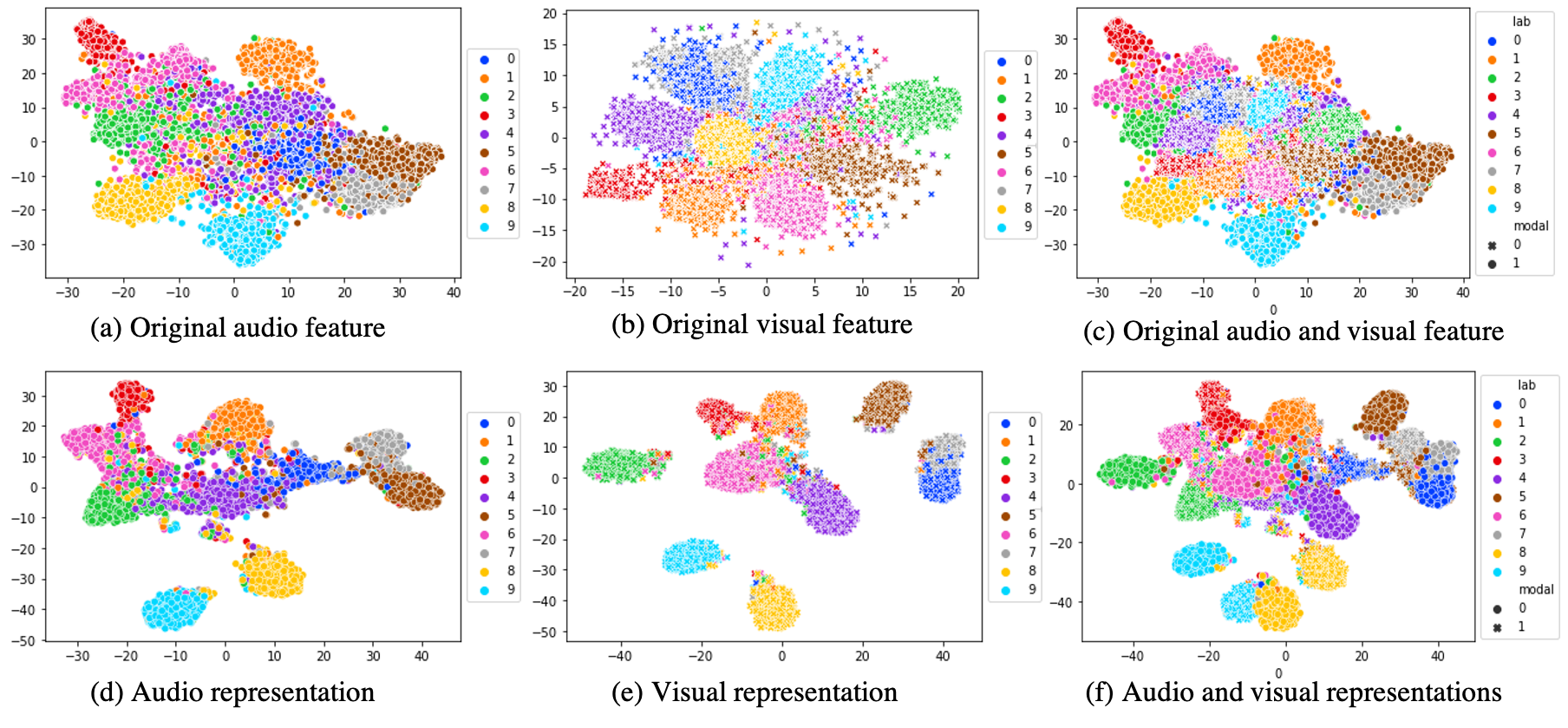}
\caption{T-SNE visualization of the test set includes original audio-visual data points and learned audio-visual representations.}
\label{fig:t_sne}
\end{figure*}

\begin{figure*}[!h]
\centering
\includegraphics[width=0.9\textwidth]{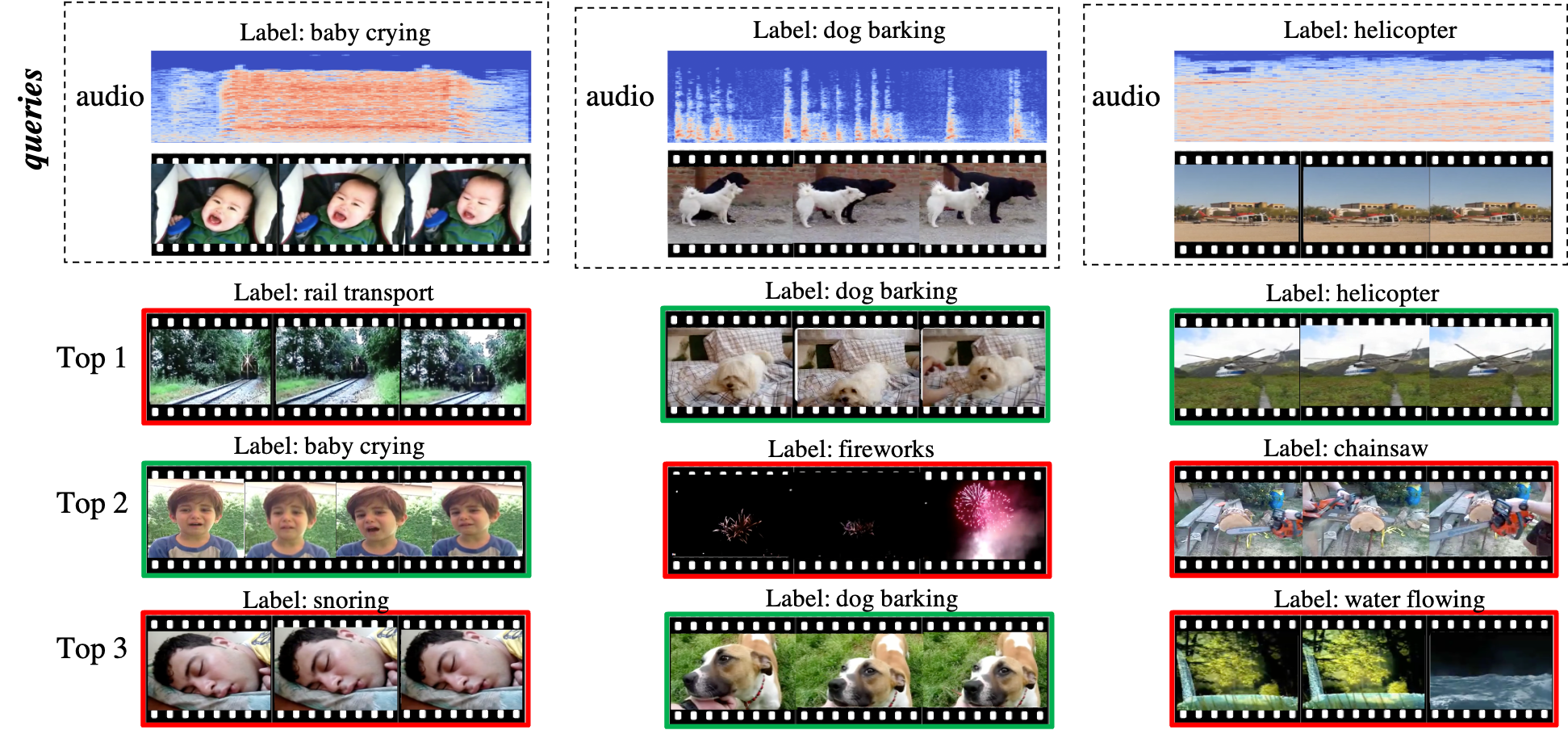}
\caption{Qualitative visual retrieved results using audio as a query on VEGAS testing set. The results are sorted from top to bottom according to the similarity score between the retrieved visual and query. The visuals in green boxes are matched with query as correct matches, and red boxes are mismatched as the wrong matches.}
\label{fig:retrieval}
\end{figure*}

\subsection{Further Analysis of our Model}
To further investigate our proposed model, we analyze our model in terms of four aspects: 1) visualization for the test data by using our model, 2) impact of the label space,  3) impact of cross-triplet combinations, and 4) audio-visual retrieval case study.

\subsubsection{Visualization of the test data using our model}
To visually leverage the effectiveness of our proposed model, we select a pair of audio and visual samples by adopting the statistical method t-SNE to reduce their high-dimensional representations into a two-dimensional space. Fig.~\ref{fig:t_sne} (a), (b), and (c) is visualizing the original 128-D audio and 1024-D visual samples by providing each sample with a location in a two-dimensional map, respectively. The distributions of original audio and visual samples of the VEGAS dataset are difficult to segregate and classify. 

Fig.~\ref{fig:t_sne} (d), (e), and (f) display the two-dimensional distributions of the audio and visual samples in the label space. We can see that the output representations across the modalities of our model are able to effectively segregates the representations into their own semantic clusters. Even though some samples are mixed together, it has been shown that our model is effective in AV-CMR in accordance with the results shown in Table ~\ref{tab:result}.

\begin{table}[h]
\caption{The MAP results of our model using label space compared with using feature common space on VEGAS dataset.}
\begin{center}
\scalebox{1.1}{%
\begin{tabular}{|c|c|c|c|}
\hline
\textbf{Model} & \textbf{\textit{Audio$\rightarrow$ Visual}}& \textbf{\textit{Visual$\rightarrow$ Audio}}& \textbf{{Average}} \\
    \hline
    Baseline-I  & 0.698 & 0.692 & 0.695\\ \hline
    Baseline-II  & 0.718 & 0.716 & 0.717\\ \hline
   \textit{\textbf{Our model} }
    & \textbf{0.766} & \textbf{0.765} & \textbf{0.766}\\ \hline
\end{tabular}}
\label{tab:labelspace}
\end{center}
\end{table}

\begin{table}[h]
\caption{The MAP results of our model compared with our baselines on the VEGAS dataset.}
\begin{center}
\scalebox{1.1}{%
\begin{tabular}{|c|c|c|c|}
\hline
\textbf{Model} & \textbf{\textit{Audio$\rightarrow$ Visual}}& \textbf{\textit{Visual$\rightarrow$ Audio}}& \textbf{{Average}} \\
    \hline
   Baseline1 & 0.175 & 0.174 & 0.175\\ \hline
   Baseline2 & 0.674 & 0.666 & 0.670\\ \hline
   Baseline3 & 0.675 & 0.663 & 0.669\\ \hline
   Baseline4 & 0.736 & 0.732 & 0.734\\ \hline
  Baseline5  & 0.679 & 0.684 & 0.681 \\ \hline
    \textit{\textbf{Our model} }  
    & \textbf{0.766} & \textbf{0.765} & \textbf{0.766}\\
\hline
\end{tabular}}
\label{tab:triplet}
\end{center}
\end{table}

\subsubsection{Impact of the label space} To bridge the modality gap that the different modality data has representations in heterogeneity, previous studies focused on projecting modality representations into the common subspace by learning pairwise correlations or other inter-modality correlation. To exploit semantic information during modality gap bridging, we take the label information to learn a label space using a linear classification-based method. We conduct some experiments to verify the advantages of improving the performance by utilizing label space instead of feature common subspace (Baseline-I and Baseline-II).

We set two baselines for the comparison experiments with the same experiment settings and shared subspaces, Baseline-I and Baseline-II use the same triplets as baseline 5 and baseline 4, respectively, in Table~\ref{tab:triplet}, instead of label space, they adopt feature common subspace where does not use the classification-based method and the dimension of representations is also set as 10. For more details of baseline 4/5, referred to subsection~\ref{sub:cross-triplet}. In Table~\ref{tab:labelspace}, by using label space, we can see our proposed model by using label space achieves better performance than the other baselines, which indicates the effectiveness of the label space learning for AV-CMR, which is consistent with Table~\ref{tab:result}.

\subsubsection{Impact of cross-triplet combinations}\label{sub:cross-triplet} To demonstrate the effectiveness of the cross-triplet function in improving the performance of AV-CMR, we set three baselines (baseline 2,3,4) that are incomplete cross-triplet losses. Baseline 1 uses the same modality to build the triplet, and baseline 5 that combines all possible triplets regardless of whether the three components are from different modalities or not. The details of the five baselines (Table~\ref{tab:triplet}) are described as follows.
\begin{itemize}
    \item Baseline 1, consists of two triplets for the ranking loss function, the components of each triplet are from the same modality so-called sing-triple. The two triplets about the modality format, i.e., (anchor, positive negative), which are: (audio, audio, audio) and (visual, visual, visual) 
    \item Baseline2, incorporates two cross-triplets, the modality format is: (audio, audio, visual)  and (visual, visual, audio).
    \item Baseline3: includes two cross-triplets, the modality format is: (audio, visual, visual)  and (visual, audio, audio).
    \item Baseline4: containing four cross-triplets, the modality format is: (audio, audio, visual), (visual, visual, audio), (audio, visual, visual)  and (visual, audio, audio). 
    \item Baseline 5: covers all possible triplets encompass single-triplets and cross-triplets, like: (audio, audio, audio), (visual, visual, visual), (audio, audio, visual), (visual, visual, audio), (audio, visual, visual), (visual, audio, audio), (audio, visual, audio), and (visual, audio, visual).
    \item Our model: comprises six cross-triplets that is a combination of baseline 2 and baseline 3, the modality format is: (audio, audio, visual), (visual, visual, audio), (audio, visual, visual), (visual, audio, audio), (audio, visual, audio), and (visual, audio, visual).
\end{itemize}
As Table~\ref{tab:triplet} shows, our model with six cross-triplets can achieve the best result compared with the other baselines. It can also be seen that the single-triplet added in cross-triplet loss can not improve the performance of AV-CMR; see the baseline 5 compared with our model. Baselines 2 and 3 as common triplet loss applied in many previous studies~\cite{wang2017adversarial, zeng2020deep, dorfer2018end} that only consider two types of triplets, which is also far fewer than our model. Overall, the results suggest that our complete cross-triplet loss in the label space is effective in improving the performance of AV-CMR.

\subsubsection{Case study}
To find qualitative results that prove our proposed model is the best of all, in Fig.~\ref{fig:retrieval}, we cherry-picked up an audio with "baby crying" event as a query to obtain its corresponding visual ranking list by using three models, i.e., the second-best model TNN-C-CCA, our baseline that uses common space for our cross-triplet loss, and our best proposed model. We only display the top five retrieved visual samples with their own labels, and based on the shortened ranking list to calculate the average precision (AP) which is AP@5 in our case. We can observe that our proposed model and TNN-C-CCA can obtain three visual samples with the "baby crying" event, our baseline can only get two visual samples with the same event as the query, so it is less effective than our proposed method. Our proposed model can achieve better AP@5 than TNN-C-CCA because the correct match that these visual samples contain the same event as the query appears further up the ranking list. Our proposed model can achieve a score of 0.7033 in terms of AP@5, outperforming the second best model TNN-C-CCA by 11.66\%. In this case study, we found that our proposed model has more accurate retrieval capabilities, identical to the conclusion drawn from Table~\ref{tab:result} and Table~\ref{tab:labelspace}.

\section{Conclusion}
In this work, we proposed a new AV-CMR model that applies label information to learn semantic discriminative features of audio-visual data in the label space and developed a cross-triplet loss to consider all the possible triplets when the candidate pairs are selected. The combination of cross-triplet loss can improve the semantic feature in the label space by capturing similar and dissimilar semantic information. We employed two different audio-visual datasets with double-check to evaluate the performance of our model with MAP and precision-scope@k evaluation metrics. The experimental results show that our model can achieve the best result compared with other state-of-the-art AV-CMR models. In the future, we would like to consider both common subspace and label space to develop a more advanced model for the AV-CMR task. 
\label{conclusion}

\section*{Acknowledgment}
This work was supported by KDDI Research, Inc. Project.

\bibliography{journal.bib}
\bibliographystyle{plain}

\end{document}